# Where does AI come from?

# A global case study across Europe, Africa, and Latin America[1]

Paola Tubaro[a]*, Antonio A. Casilli[b], Maxime Cornet[b], Clément Le Ludec[c], Juana Torres Cierpe[d]

[a] *Centre of Research in Economics and Statistics (CREST), Centre National de la Recherche Scientifique (CNRS), Institut Polytechnique de Paris, 91120 Palaiseau, France*

[b] *SES Department, Interdisciplinary Institute for Innovation (i3), Telecom Paris / Institut Polytechnique de Paris, 91120 Palaiseau, France*

[c] *CERSA, Université Paris-Panthéon-Assas, 75005 Paris, France*

[d] *LaborIA, Inria, 78150 Le Chesnay-Rocquencourt, France*

*\* Corresponding author: CREST-ENSAE, 5 avenue Henry Le Chatelier, 91120 Palaiseau Cedex, France, Tél. +33 17026 6883, paola.tubaro@ensae.fr*

**Abstract**

This article examines the organisational and geographical forces that shape the supply chains of artificial intelligence (AI) through outsourced and offshored data work. Bridging sociological theories of relational inequalities and embeddedness with critical approaches to Global Value Chains, we conduct a global case study of the digitally enabled organisation of data work in France, Madagascar, and Venezuela. The AI supply chains procure data work via a mix of arm's length contracts through marketplace-like platforms, and of embedded firm-like structures that offer greater stability but less

---

flexibility, with multiple intermediate arrangements. Each solution suits specific types and purposes of data work in AI preparation, verification, and impersonation. While all forms reproduce well-known patterns of exclusion that harm externalised workers especially in the Global South, disadvantage manifests unevenly in different supply chain structures, with repercussions on remunerations, job security and working conditions. Unveiling these processes of contemporary technology development provides insights into possible policy implications.

**Funding details**

This work was supported by Force Ouvrière (FO) with Institut des Recherches Economiques et Sociales (IRES) under Grant AO2017; Maison des Sciences de l'Homme (MSH) Paris-Saclay under Grants 17-MA-07 and 20-MA-02; Centre National de la Recherche Scientifique – Mission pour les Initiatives Transverses et Interdisciplinaires (CNRS-MITI) through 'Enjeux Sociaux et Scientifiques de l'IA 2020'; and Agence Nationale de la Recherche (ANR) under Grant ANR-19-CE10-0012.

**Disclosure statement**

The authors report there are no competing interests to declare.

**Introduction**

While the current hype around artificial intelligence (AI) spells disruptions to labour markets and the workplace, calling policymakers to action (OECD 2024), more limited attention is devoted to the labour inputs that are needed to produce AI itself. Indeed, beyond high-earning engineers and computer scientists who design 'smart' solutions, myriad low-level workers prepare data to train and test models, perform quality checks, and make corrections when algorithms malfunction. A landmark example is the ImageNet database for computer vision, through which so-called 'Deep Learning' revealed its power and potential to the world in 2012, and whose development required over 50,000 people behind the scenes to manually label its 14 million images (Denton et al. 2021).

The early literature (for example, Ross et al. 2010) stressed how this tech-driven 'data work', also called 'microwork' or 'crowdwork' (Muldoon et al., 2024), was procured through digital labour platforms, using technology to match demand and supply for online gigs without any long-term commitments. ImageNet workers were recruited, managed and paid through Amazon Mechanical Turk, a pioneer of this format since the mid-2000s, with a marketplace-like design allowing scale and speed while keeping costs down. In such a setting, labour contracts take the form of arm's length transactions between clients (tech companies) and users (workers), both construed as independent businesses. The platform-as-marketplace model escapes the rules of salaried employment and raises concerns about erosion of job security and social protection, just like 'uberised' digital labour in ride-hailing and delivery services (ILO 2021).

Recent evidence points to greater diversity in the labour supply chains behind AI. Schmidt (2019, 2022) notes a shift from 'general-purpose' to 'specialised, full-service crowd-AI stack' platforms that do not function as pure online marketplaces. There are also cases of brick-and-mortar firms, with internal hierarchies and employment contracts (Miceli et al. 2020, Miceli and Posada 2022). Some are subsidiaries of, or specialized offices within, multinational companies (Muldoon et al. 2023). Others, like the 'follower factories' observed by Lindquist (2022), are part of the informal economy. Alongside the Mechanical Turk model, corporate-like organisations resurface and bring AI into the broader, classical debate around modes of economic coordination between the two poles of market and firm (Coase 1937). This tension has become part of the definition of a digital platform as a hybrid form, which also meets three other criteria: running a software-as-a-service (SaaS) infrastructure to put users to work, 'taskifying' labour processes to generate data, and striving toward automation (Casilli and Posada 2019).

Any attempts to map the current diversity of organisational structures behind AI production, and the factors driving them, would be incomplete without considering that data work has gone global. First documented in the United States where three quarters of Mechanical Turk workers live (Difallah et al. 2018), data work is now also present in Europe (Morgan et al. 2023), and it has especially expanded into the Global South (Datta et al. 2023) with hubs in India (Gray and Suri 2019), Brazil (Grohmann and Fernandes Araújo 2021), Argentina (Miceli et al. 2020), Venezuela (Posada 2022), Kenya (Muldoon et al. 2023) and other parts of sub-Saharan Africa (Anwar and Graham 2022). In these places, a variety of organisational forms have been observed, but the reasons of these differences and the ways data work dovetails with local economies remain under-researched.

The focus of this article is the way platforms interact with the international supply chains of AI in different regional settings. It clarifies the structure and organization of these supply chains, highlighting their impacts on labour conditions and remunerations. At the crossroads of relational inequalities and embeddedness theories in economic sociology, and critical approaches to Global Value Chains (GVC) in political economy, we conduct within-case and cross-case analyses of three in-depth case studies across the planet. Results show that the AI supply chain is best understood as an instance of the outsourcing and offshoring trends already observed in other globalised industries. While AI is commonly depicted as a disruptive force capable of profoundly reshaping contemporary economic life, the value chains that underpin its production reproduce well-known patterns of exclusion that hurt growing numbers of externalised workers and widen the gap between the Global North and South. However, disadvantage takes multiple forms as different types of data work are associated with distinct organizational structures and geographies, with repercussions on wages, job security and working conditions along the supply chain. Thus, responses based only on worker re-classification are one-sided, and we advocate a policy mix with combinations of tools adapted to local conditions, at both national and supra-national levels.

**Theoretical background and research questions**

*Unveiling the role of data work in the AI supply chain*

The contribution of data work to AI production was established early (Irani 2015) and repeatedly confirmed (Gray and Suri 2019, Crawford 2021). Contemporary AI relies on human inputs to generate and annotate data that feed machine learning models. For example, recorded human utterances are the first ingredient to develop voice assistants like Alexa and Siri (data generation). These recordings then need to be transcribed and labelled, indicating for example voice tone, speaker's sentiment, and characteristics like gender and age (data annotation). Tubaro et al. (2020) define these activities as 'AI preparation' and note that there are two more human inputs needed at later stages, namely 'AI verification' and 'AI impersonation'. Verification is the process by which output accuracy is checked once the AI solution has been released: for example, manual examination of search engine results to determine whether they correspond to users' queries. Impersonation occurs when humans act fully or partly in place of unreliable algorithms, although their presence and role are often concealed behind an overwhelming rhetoric of automation – which Rani and Kumar Dhir (2024) call 'deceptive AI'.

This analysis has two simple, but far-reaching implications. One is that data work encompasses lower-level functions. Data tasks are typically described as simple, short, and repetitive. Although recent studies reveal the emergence of more complex and lengthy tasks (Schmidt 2022), their added value remains limited. The other is that data work produces intermediate inputs for AI production, not goods for final consumption; its organisation will therefore largely depend on the strategic choices of the technology and research organisations that use it in their productive processes. Let us first derive the direct consequences of these properties, before discussing the extent to which the three functions of data work involve different organisational solutions.

*Outsourcing: A barrier for data workers' inclusion*

Data work contributes to AI manufacturing outside the boundaries of technology companies and research centres. The producers of chatbots, search engines and self-driving cars do not usually employ lower-level, low-paid data workers but rather outsource these services to external providers. As already observed in multiple industries, these companies focus on core areas and shed peripheral activities to outside contractors, suppliers, and subsidiaries (Weil 2014). By shrinking their boundaries, high-earning tech firms avoid any responsibility to manage, supervise, and set wages for lower-level workers, while still benefiting from their productive efforts and, in many cases, imposing and enforcing strict standards to frame their activity.

While outsourcing strategies can be highly profitable for the lead company, relational inequality theory (Tomaskovic-Devey and Avent-Holt 2019) highlights that they disadvantage externalised workers. As outsiders, they contribute to core company profits but have limited bargaining power and cannot make claims on those profits. They typically operate in highly competitive environments that put downward pressure on their remunerations, while they also face the cost of meeting the (sometimes demanding) goals and standards imposed on them. Thus, remunerations and working conditions vary across the boundaries of the lead firm, irrespective of any differentials in skills and experience.

Without externalisation by technology producers, the data work market and its diverse intermediaries would not even exist. The structure of the AI supply chain results from exclusion of low-level functions from the wealth of financial, cognitive, and reputational resources that technology-producing companies attract today. Whether data work is procured through marketplace-like digital platforms à la Mechanical Turk or seemingly conventional firms makes little difference, except that the former allow outsourcing to formally independent micro-providers instead of multi-person operations (Lehdonvirta et al. 2019). Either way, the lead AI producer imposes standards and priorities. Thus, we can expect low remunerations and poor working conditions for all data workers, regardless of their legal status or classification.

*Offshoring: Lower cost of data work in the Global South*

As noted above, data work is geographically spread, while higher-paying functions such as branding and product design remain largely concentrated in the Global North. Because data work can be executed remotely, it can be viewed as a tradable international service, part of a planetary market (Graham and Ferrari 2022). The GVC framework (Gereffi et al. 2005) explains how outsourcing can combine with offshoring: like other industries, AI leverages recent advances in telecommunications and transportation to gain competitive advantage.

GVC analysis frames companies' offshoring decisions in terms of efficiency and transaction costs, potentially allowing poorer countries to progress up the supply chain. Less optimistically, Milberg and Winkler (2013), highlight how core companies in richer countries intentionally maintain strong market power asymmetries that increase their surplus and prevent their contractors from upgrading, while maintaining low costs for imports (including labour costs). As mono- or oligopsonistic buyers of intermediate inputs, they foster competition between their suppliers and create excess capacity, thereby keeping remunerations low.

Likewise, Selwyn (2016, 2019) contends that cross-country wage gaps do not arise from lower productivity in the Global South, but from firms in the North exploiting or 'super-exploiting' (paying below subsistence level) workers. He introduces the concept of 'Global Poverty Chains' to illustrate how these practices fail to integrate offshore workers and to improve their living conditions. These conclusions resonate with evidence that the management systems of data work platforms constrain skill development (Rani and Furrer 2019), so that low productivity may be a consequence rather than a cause of low pay. For the purposes of this article, we thus anticipate that AI producers' value-capturing strategies and market power will reduce remunerations, deteriorate working conditions, and slow productivity growth, particularly when data work is offshored to the Global South.

*The embeddedness of AI supply chains*

How do AI companies choose between one-person micro-businesses accessed through a marketplace platform like Mechanical Turk, and firm-like vendors? Economic sociology suggests that the relationships of a lead firm with its suppliers and contractors are not always impersonal and profit-oriented arm's length contracts, where actors look for the best prices at any moment in time. Some commercial ties are, instead, embedded in longer-term personal relationships that enable thick information exchange, shared risk management, and mutual learning. If arm's length commercial ties, that is, transactions that exclude any prior connection, allow flexibility (for example, capacity to quickly increase production in response to demand peaks), embedded ties help when trust is needed to solve joint problems. Because they perform different functions, multiple types of ties often coexist within the same production network (Uzzi 1996).

Originally meant to describe business leaders, these analyses have implications for workers that can be derived from the redefinition of embeddedness of Brailly et al. (2016) as the combination of 'vertical' affiliation ties between employers and workers, and 'horizontal' ties between peers within each group. Involved in both types of ties, embedded workers likely benefit from better access to information, a more predictable workflow, and clearer rules. Data work offers an excellent setting to apply these ideas. Amazon Mechanical Turk and other marketplace platforms enable the creation of arm's length ties in which workers are largely disembedded, but firm-like arrangements with human management are all based on some degree of trust between the contractor and the lead company, suggesting embeddedness. As with the multi-million-item ImageNet dataset, marketplace-managing platforms allow scale, but their sheer size and anonymity hamper monitoring and control. Researchers (Schmidt 2019) and industry sources (Cognilytica 2022) agree that confidential and sensitive data as well as intellectual property protection discourage use of Mechanical-Turk-like digital platforms, especially among companies developing proprietary AI. Today, even platforms that were originally designed as self-service marketplaces, such as Clickworker and Appen, increasingly offer managed service solutions that not only hire, but also train and supervise groups of workers for specific projects, shifting the balance from the market toward the firm as a reference model.

The remainder of this article sets out to identify whether the three main functions of data work – AI preparation, verification, and impersonation – correspond to different degrees of embeddedness. The above considerations suggest that arm's length ties are appropriate when quality matters less than volume, or to meet occasional needs like peaks in demand. We expect this to happen more often with AI preparation, especially data generation. Impersonation and verification are generally carried out at smaller scale but to higher quality standards, thus benefiting from embedded ties. Depending on the data tasks they perform, workers will therefore be part of differently organized supply chains, resulting in unequal working conditions.

**Methods and data: Finding patterns in AI supply chains across three continents**

The breadth and the innovative nature of these questions indicate that the most appropriate methodology is a qualitative, albeit large-scale empirical approach. A study of data work in France (2018-19) initiated our efforts to document the scope and specificities of this emerging phenomenon outside the English-speaking world, already researched since about 2010 (Ross et al. 2010). We expected the supply chains of data work to (partly) form along linguistic lines, to feed language-based applications of AI (for example voice recognition, automated translation, and text generation). Our study quickly revealed that many French tech companies seek data workers not only locally, but also in low-income, mostly francophone countries with commercial and cultural ties to their previous motherland.

Therefore, in 2020-22 we expanded our fieldwork into Madagascar. A former colony, where 26% of the population speak French, often as a second language (OIF 2022), the African island nation is a net exporter of computing services, with a growth of 65% between 2005 and 2019 (WTO 2022). Among African French-speaking countries, Madagascar ranked second in terms of computing service exports (73 million dollars) in 2018, and foreign investors benefited from a favourable free zone regime. Another African country, Egypt, offers a point of comparison, with its larger, but less export-oriented computer services sector, a small French-speaking population (3%), and a diverse set of commercial relationships to Europe and Asia.

Meanwhile, we extended our fieldwork into Spain and Spanish-speaking Latin America (2020-22) as examples of another widely used language other than English. As a European country, Spain allows comparison with France, while its historical ties to Latin America make it a benchmark for the whole Spanish-speaking world. Among these countries, Venezuela is a particularly interesting case insofar as the severe economic and political crisis that has plagued it since the mid-2010s has prompted many people to work remotely on data work platforms (Schmidt 2019). We use the other Central and South American countries (Argentina, Bolivia, Chile, Colombia, Costa Rica, Ecuador, Dominican Republic, El Salvador, Guatemala, Honduras, Mexico, Nicaragua, Panama, Paraguay, Peru, and Uruguay) as comparison points.

We adopt a case study approach, which involves an all-inclusive, in-depth analysis of each of the main settings (France, Madagascar, Venezuela) in relation to its comparison points. Unlike experimental designs, case study research does not impose controlled conditions and is therefore less suited to testing causal hypotheses, but it sheds light on the role of the real-life contexts in which the phenomena of interest occur. It is helpful for disentangling multifaceted and complex phenomena, especially when there is scarce prior knowledge of the population or background (Yin 2017). Unlike static methods, a case study allows close investigation of dynamic social processes as they evolve. To remain as open as possible, to demonstrate that the phenomenon is global, and to bring evidence that it is not a one-off occurrence, we have chosen widely differing settings in terms of geography and, as we will see, of organizational arrangements.

For each case, we collected multiple sources of evidence, mixing quantitative and qualitative techniques, and relying on data triangulation to ensure consistency of findings. In France, we collected 908 questionnaires from data workers of a local platform – well known nationally and only recruiting residents – along with 92 interviews with data workers, tech companies, and data service providers. Local platforms are major, albeit under-researched, actors in these markets (Datta et al. 2023). Data workers in Madagascar filled out 294 questionnaires and we interviewed 215 stakeholders, 180 of whom are workers; we also conducted observations at four data companies in the capital, Antananarivo. In Cairo, Egypt, we interviewed managers of three data work companies. Spanish and Latin American data workers from two large international platforms filled out 2428 questionnaires, and 58 of them also participated in in-depth interviews. The studies were approved by the Data Protection Officers of our institutions, namely of Telecom Paris (France, Madagascar, Egypt), and of CNRS (Spain and Latin America). Informed consent was obtained from all participants.

The diversity of types, sizes, and varieties of data gathered is both a manageable limitation and a valuable asset for our research. In the Spanish-language study, workers' strong internet presence allowed us to field questionnaires online and to do interviews via videoconferencing tools. But these techniques would be ill-suited to study companies with a physical location, so we conducted the Egypt fieldwork in-person (although COVID-19 restrictions downsized our original plans), and most of the Madagascar study also in-person after a preliminary online phase. These discrepancies prevent cross-country comparisons strictly speaking, and we use instead a 'collective' case study approach (Stake 1995), albeit on a global scale. In each setting, we combine the sources available to illuminate the specific aspect of our research problem that it best exemplifies, aiming at depth and concreteness of analysis; we then juxtapose the different cases to generate a broader understanding.

*Europe (France)*

It was in France that we first saw how different organizational arrangements supply data work to tech companies. We recruited most study participants on a local platform that launched itself as the French equivalent of Amazon Mechanical Turk, but then shifted from this marketplace-oriented toward a managed-service model. We also interviewed users of international data work platforms.

Both local and international platform marketplaces offer primarily, though not exclusively, AI preparation tasks. Recording one's voice, anonymising CVs, and tagging objects in images were part of the experience of many surveyed data workers. Combinations of firms and platforms, sometimes in convoluted configurations, handle verification tasks, which are often complex or sensitive. In one case, a French worker who was tasked with checking whether a voice assistant correctly understood user requests, served a US-based lead firm (a tech company), through a Chinese platform which had availed itself of a Japanese website to post the job and of a Spanish subcontractor to undertake the recruitment, while another contractor in Italy ensured day-to-day management.

Outsourcing occurs in France along with limited or no offshoring. The platform where we recruited participants is open only to French workers, and clients are also mostly French. International platforms, like those used by Spanish study participants, open tasks to anyone willing to do them, or sometimes reserve them to given countries when local knowledge or language are needed.

The vast majority of French respondents work from home, often without any relationship with peers, clients, or intermediaries. They leverage the excellent connectivity infrastructure of the country (in 2021, 93 % of households had internet access in France, INSEE 2022), modern computing equipment, and in many cases university-level education. Slightly over two thirds of workers engage in data tasks in parallel to another job elsewhere, often to raise extra income (for example when the other job is part-time), or intermittently (after a seasonal or temporary job). They rarely undertake these activities full-time, largely because remunerations are too low and volatile compared to the country's cost of living. Spain reveals a similar picture, although with a few more people practicing data work full time as a buffer between jobs. As a result, most workers feel disengaged and unwilling to devote time and energy to learning or sharing information with others, not to mention organising to make claims on the product of their work.

*Africa (Madagascar)*

One-person micro-providers who perform data work from home through online platform marketplaces are more sporadic in Madagascar because of the high costs of internet connections. Small and medium computing companies, often with physical offices, employ local workers to fulfil orders from (mainly) French clients that demand large amounts of data for longer projects or impose specific confidentiality and quality requirements. Likewise in Egypt, several small firms meet data-work demand from companies in China and the Gulf. These organisations follow in the footsteps of those that have supplied overseas businesses since the 1990s, like call centres. Already established in this industry, Madagascar and Egypt have added data services to their offer for clients in France and elsewhere.

Platforms like Mechanical Turk take a less prominent role without disappearing from the picture. Some of these local companies internalise their use, harnessing only their technical facilities (like image annotation software). Other companies arise informally when highly reputed workers secure tasks on international marketplace platforms and redistribute work to relatives and friends without officially employing them, setting up small operations in their living rooms or even in cybercafes.

The financial health of even formal firms is often precarious, as demonstrated by crowded suburban offices and heavy workloads for staff members. Therefore, these arrangements offer limited advantages. Malagasy workers have employment contracts with local firms, access to office facilities (including computers and internet connection), and daily contact with colleagues, managers, and sometimes clients. Pay is comparable to other jobs in the local outsourcing sector, and relatively stable insofar as negotiation of large deals offers contractors a minimum of bargaining power. However, contracts are often short-term owing to fluctuations in foreign demand, payment delays are common, and real wages lag behind inflation. These data workers are located at the end of long outsourcing/offshoring chains that exclude them from the resources that the technology industry accumulates in high-income countries.

We observed AI verification only in formal data companies, and AI preparation and impersonation in all. Data entry for e-commerce websites and transcription of newspaper articles are among the preparation tasks that Malagasy workers do, along with map updating for military purposes—which requires access to sensitive data. Our investigation uncovered an egregious case of impersonation involving 'smart' surveillance cameras deployed in some French retail chains and supposed to automatically alert security agents in cases of suspected shoplifting. In fact, the detection was carried out manually by workers in the outskirts of Antananarivo (Le Ludec et al. 2023). According to one data company owner interviewed in Egypt, workers based in a suburban Cairo office check Chinese face recognition camera feeds, although we could not ascertain whether this was also a case of AI impersonation.

Founders and managers of these companies insist on the quality they achieve, compared to pure marketplace platforms. To meet their clients' standards, they implement training, supervision, and controls. Their workforce is small, but highly qualified. In Madagascar, it includes recent graduates (mostly men), often proficient in French, who hope to get a better job after gaining experience with data tasks. Low pay discourages workers with family responsibilities in increasingly expensive Antananarivo. Egyptian data preparation companies also hire university-level data workers, but in some cases, they are mostly young unmarried women, who get engaged for two or three years before starting a family or moving elsewhere.

*Latin America (Venezuela)*

Although firm-like vendors exist in some Latin American countries like Argentina (Miceli et al. 2020), it is on online marketplaces managed by international platforms that the relevance of Venezuela in the data work supply chains has become apparent. Partly because it could thrive on oil until the mid-2010s, partly for political reasons, the country has no tradition of providing outsourced computing services, which constituted only 0.02% of its exports in 2005-10 (UNCTAD 2022). The collapse of the bolivar and the daily price changes that make even staple foods unaffordable to those earning local wages have prompted many highly educated Venezuelans to embark on platform work as a pathway to international labour markets and to payments in hard currency (Johnston 2022).

Like French data workers, Venezuelans are single-person micro-providers; like Malagasy workers, they are part of global offshoring chains that supply AI inputs to the Global North. But unlike the other two, Venezuelans occupy the lowest-paid segments of the market. Failing infrastructure excludes them from tasks that require higher computing power or fast connection, and they often perform very short and poorly paid 'clickfarm' tasks that consist of liking, sharing, and reviewing online content. These tasks serve immediate marketing goals, but they also feed into AI preparation. The low value of the domestic currency explains Venezuelans' willingness to do the lowest-paying tasks, such as captcha-solving, usually at rates of 0.50 – 1.50 USD per one thousand items. Supposedly meant to support cybersecurity, these tasks can be used maliciously, if an intruder hires human workers to solve the captchas that protect a website, then attacks it with bots. Even when tasks are better paid, local conditions push Venezuelan data workers to accept the riskiest among them: for example, taking selfies to build face recognition datasets, at 3 – 5 USD per set, but with anxiety about the potential reuses of one's own image.

Though largely prevalent, the marketplace model does not exhaust the varieties of organisational arrangements in place. To convert and withdraw the currency earned online, Venezuelan data workers go through several layers of online financial intermediaries, be they formal companies, individual brokers, or black-market actors. A similar, albeit less extreme, picture is observed in Argentina. The resulting intricate structures are not without reminding those observed in France but come with the additional burden of transaction fees and financial risk (Posada 2024).

Low cost does not always make Venezuelan data workers attractive to clients. When our interviews took place, one platform had introduced a 'quota' giving them access to only 10% of all tasks within international data projects. This measure partly aimed at ensuring diversity of workers, and partly at controlling performance standards, to combat allegedly 'unsatisfactory' Venezuelan work. Though temporary, the quota exacerbated competition and evicted less efficient workers – those who failed to submit their outputs quickly enough, before all tasks in a project were completed – in ways we have not seen in any other Latin American country. This episode showcases how client companies and platforms use perceived poor quality to exclude some workers and to pay the others less. Interviewed workers were aware of the bad reputation of their country on platforms, and many believed that for the same tasks, they were being remunerated less than their peers in, say, Colombia. The evidence we collected suggests that such quality issues are rather due to weak infrastructure: unstable electricity supply, slow internet, and outdated computers.

Although Venezuelan data workers typically perform simple and repetitive tasks, they often demonstrate creativity and know-how: automating recurring tasks via hotkeys, writing scripts to get around the infamous quota, using virtual private networks (VPNs) to spoof location... Some of these solutions violate platforms' terms of service and may result in exclusion and loss of earnings if discovered. But they help to compete on international platform marketplaces, and workers share them enthusiastically. In the country, platform data work has become a common undertaking among family members, friends, and neighbours. On social media like Facebook, WhatsApp, and Telegram, Venezuelans exchange information, advice, and mutual support. They are more dependent on platform data work, and less socially isolated, than any other Latin American country in our sample.

**Results: Platforms as coordination devices that structure AI supply chains**

These case studies illustrate various instances of outsourcing, whereby data work is undertaken outside the boundaries of the organisation that produces AI technologies. Invariably, they take place on digital platforms as defined by Casilli and Posada (2019), with a technical infrastructure that generates data through users' taskified work and serves automation purposes. The difference is that in some cases, platforms take the form of completely remote work infrastructures framed as online marketplaces that match clients and one-person micro-providers (workers), while in others, they are firms that manage teams of workers under their hierarchical structure. Both these extremes qualify as digital platforms, but manifest differently, with a continuum of organisational arrangements between them. A particularly interesting intermediate case is the chain of intermediaries that we first observed in France, and that resurface (to a lesser extent) in Venezuela for payment management, where multiple layers of contractors, brokers, and suppliers frame workers' activity in structures that are evocatively reminiscent of complex neural networks in so-called 'deep learning' (Schmidt 2022, Tubaro 2021, Posada 2024) and that we therefore relabel 'deep labour'.

Organisation of data work along the marketplace/firm platform continuum leads to different forms of employment and worker status. Marketplace platforms establish arm's length relationships between data workers and clients. This does not mean only absence of long-term commitments – namely employment relationships – but also paucity of interactions and contacts. In France, this implies isolation from peers, which leads data workers to feel alone and, perhaps more importantly, to lack opportunities to learn from one another. The experience of Latin American workers is similar, though less so in Venezuela, where workers rely on their close contacts for mutual support. In both online and offline environments, they harness their personal relationships to attenuate the negative effects of disembeddedness and to overcome technical difficulties. At the other extreme, firm-like organisations in Egypt and Madagascar typically form embedded ties with their overseas clients, for projects that can last up to a few years. Data workers who are trusted by their local contractor and their overseas lead firm enjoy less volatile remuneration as well as opportunities for socialisation, learning, and (limited) career progression. Deep labour platforms fall somewhere in-between, combining arm's length and embedded ties. Although they do not employ workers strictly speaking, they provide training, monitoring, and support services.

Regardless of organisational structures, data workers' rights are always adversely affected by their exclusion from the wealth and resources of AI-producing companies. Lack of employment contracts keeps French platform data workers separate from client companies, while low remunerations discourage strong commitment and with it, any sense of entitlement to those companies' surplus. Even workers with employment contracts, like those in Madagascar, face precarious working conditions insofar as their relationship is not with the lead companies that buy the product of their work to develop new technologies, but with local firms endowed with limited financial resources and under pressure from competitors in uncertain environments. Power gaps between data workers and platform clients are at their highest in Venezuela, where geographic and symbolic distance from the lead company compounds with highly volatile arm's length contracts and the wage-depressing effects of extremely tight competition. Marketplace-like platforms overwhelmingly favour clients, and any claim-making by workers (even when it is only to report a technical bug) becomes so burdensome that they sometimes give up without trying.

As we have seen, data work is a global phenomenon at the intersection of outsourcing and offshoring: data workers in Africa and Latin America mostly serve overseas companies. Lower remunerations in Global South countries do not necessarily mean lower productivity but reflect lack of recognition of activities that are not meant to be disclosed. The Madagascar AI impersonation case illustrates how, confused by corporate narratives of full automation, society fails to appreciate the value and utility of manual data processing. The mispricing of data work sustains the aura of AI and its supposedly unstoppable progress. Other factors like excess supply and fierce competition among data task providers, particularly acute in Venezuela, lower pay directly and indirectly, through the negative impact they have on quality (hence, productivity) when combined with insufficient equipment and/or unfavourable macroeconomic backdrop.

**Discussion: Where do AI preparation, verification, and impersonation happen?**

To understand how firms choose between these different organizational arrangements, we must look at the tasks that data workers perform. With its emphasis on 'raw data', AI preparation can be compared to exports of semi-finished products, while as a way to ensure that marketed services function correctly, AI verification can be seen as a form of customer support. AI impersonation is occasionally driven by malicious intent, but it can also constitute the first step in machine learning pipelines, and it may contribute to troubleshooting AI systems.

Figure 1 summarizes our main findings. AI preparation is widespread on marketplace platforms, although it also occurs in other supply chain structures. In most cases, data entry and data generation are considered low-value, standardised activities, which do not require data workers to be familiar with the client's strategic orientations and patented technologies. Arm's length ties are sufficient, as evidenced by the fact that these activities are observed in all three settings. In contrast, when workers need to gain access to confidential information or acquire a deeper understanding of a proprietary technology to perform data tasks, the situation changes dramatically. Trust becomes paramount, and embedded ties gain importance. This issue is especially sensitive in the case of AI impersonation activities, which risk to be interpreted as fraudulent activities, technical bugs, or system failures. These situations call for the backing of firm-like platforms that give workers more secure contracts in return for discretion—as in Madagascar—or of deep labour platforms that work with other suppliers to mitigate liability and ensure a stable yet flexible workforce—as in France.

[INSERT FIGURE 1 HERE]

Data quality is another factor frequently used by companies involved in AI production to justify their managerial decisions. Marketplace platforms invoke poor data quality as a reason to establish only a limited amount of arm's-length contracts with Venezuelan workers. On the opposite side of the spectrum, European AI companies justify their use of embedded and relatively stable contracts with Malagasy contractors to ensure higher quality data work. It seems that platform organisational structures affect labour quality and job stability. But quality is defined differently depending on the type of data work performed: AI preparation, especially data generation tasks, requires diverse data sources and the ability to scale quickly, making it more compatible with marketplace platforms. In this setting, data quality is synonymous with variety and volume. In contrast, AI verification and impersonation involve more control and compliance with lead companies' standards. Stable work environments reduce competition between workers, encourage cooperation and allow more complex tasks, deemed to be more lucrative. This scenario is best suited to firm-like platforms, where quality comes with accuracy, teamwork, and reliability.

*Conclusions*

Bridging relational inequality theory, critical approaches to GVC, and the sociology of embeddedness, our investigation highlights continuity rather than disruption. The AI supply chain mirrors the patterns of outsourcing and offshoring already observed in domains like consumer electronics and textiles. Although the future impacts of potential massive deployment of AI in the economy remain to be established (particularly in respect to the expectation that automation may destroy jobs), its very production, at today's still early stage, perpetuates a well-known trend toward rising inequality between labour and capital. Importantly, against a literature on data work which has commonly assumed that disadvantage ensues from absence of employment relationships, the analysis conducted above turns the problem on its head by showing that it is a by-product of exclusion from the resources of lead AI producers, and therefore hurts all data workers however classified. The narrow focus of earlier studies on Amazon Mechanical Turk and similar marketplace platforms prevented a thorough understanding of the landscape of human supply chains for AI.

If disadvantage affects everyone, it unfolds differently depending on geography – itself a proxy for institutional, cultural and linguistic proximity – and degree of embeddedness. In the countries where AI companies are based, workers have access to better-paid tasks and alternatives to fall back on, which make them less dependent on data tasks for living. In firm-like platforms that negotiate larger-sized or longer-term contracts, workers enjoy less volatile remuneration and greater job security. Our three cases are differentially positioned along these two analytical axes with, so to speak, France being geographically close but (often, though not always) disembedded, Madagascar distant but embedded, and Venezuela both distant and (largely) disembedded.

Organizational arrangements currently exhibit a great deal of variation, as companies choose diverse solutions along the continuum between the marketplace and the firm, such as those that we have labelled deep labour. The configurations observed today may not be stable but represent a snapshot in a longer evolution whose ultimate outcomes will depend on future technological developments, market trends, and regulatory responses. Our case studies uncover key factors informing business decisions, but they cannot establish whether tech companies will continue to move freely across the organisational spectrum or will eventually converge toward one of the poles.

Although we have emphasized continuity, we recognise that digital technology, through the pervasiveness of platforms, has transformed outsourcing and offshoring in all observed arrangements. This change is best understood in reference to the above-cited work of Casilli and Posada (2019). Based on the results of our investigation into how platforms impact supply chains, the first criterion in their definition (a hybrid between a firm and a marketplace) should be tempered or rephrased ('platforms can be firms, marketplaces or anything in between'), while the other three remain powerfully in place: these organisations use powerful software architectures to have workers perform fragmented data tasks, in order to develop AI systems. Data production, extreme decomposition of labour processes, and automation define and make possible this very market and its present configurations.

This analysis has obvious limitations, primarily stemming from heterogeneity of the data sources used, focus only on a selected number of country settings, and need to 'connect the dots' without performing a full-fledged comparative analysis. Still, it has shed new light on the AI industry and its impacts on labour, globalisation, and development. Its focus on the accumulation strategies of lead firms contributes to a better understanding of the changing power relations between capital and labour in this rapidly evolving field. Our work also suggests important policy implications. Especially in Global North countries, efforts are being undertaken to improve the fate of platform workers via reclassification. Though in many respects laudable, these initiatives fail to address the broader range of issues that arise with AI data work and the variety of organizational structures that frame it, particularly in Global South countries. Our research indicates a need for an appropriate, ideally supra-national policy mix that in addition to changes in labour law, also envisages appropriate regulation of technology and innovation, and promotes suitable strategies for economic development. Such mix could include due diligence legislation as recently promulgated in France, requiring lead firms to take responsibility for respect of labour legislation and human rights all along their supply chains, even beyond corporate boundaries and national borders. Initially intended for large multinationals in general and not for technology companies specifically, such laws may need some adaptation to account for sectoral specificities (for example, to ensure applicability to smaller AI start-ups) but can provide a powerful legal tool to monitor and improve working conditions throughout the AI supply chain.

**Acknowledgements**

We thank three anonymous reviewers and the editors of the Special Issue for their feedback on preliminary versions of this article.

**Figure 1. Marketplace- and firm-like platforms in the supply chains for data work in Europe, Africa, and Latin America.** Dark grey countries: main case studies, light grey countries: comparison cases. Organisational modes range from almost totally marketplace oriented (darker rectangle, Venezuela) to almost entirely firm oriented (lighter rectangle, Madagascar). AI preparation (darker circle) is ubiquitous, but AI verification (darker triangle) and AI impersonation (darker star) tend to happen in 'deep labour' and firm-like organisations where embeddedness is higher. Source: authors' elaboration.

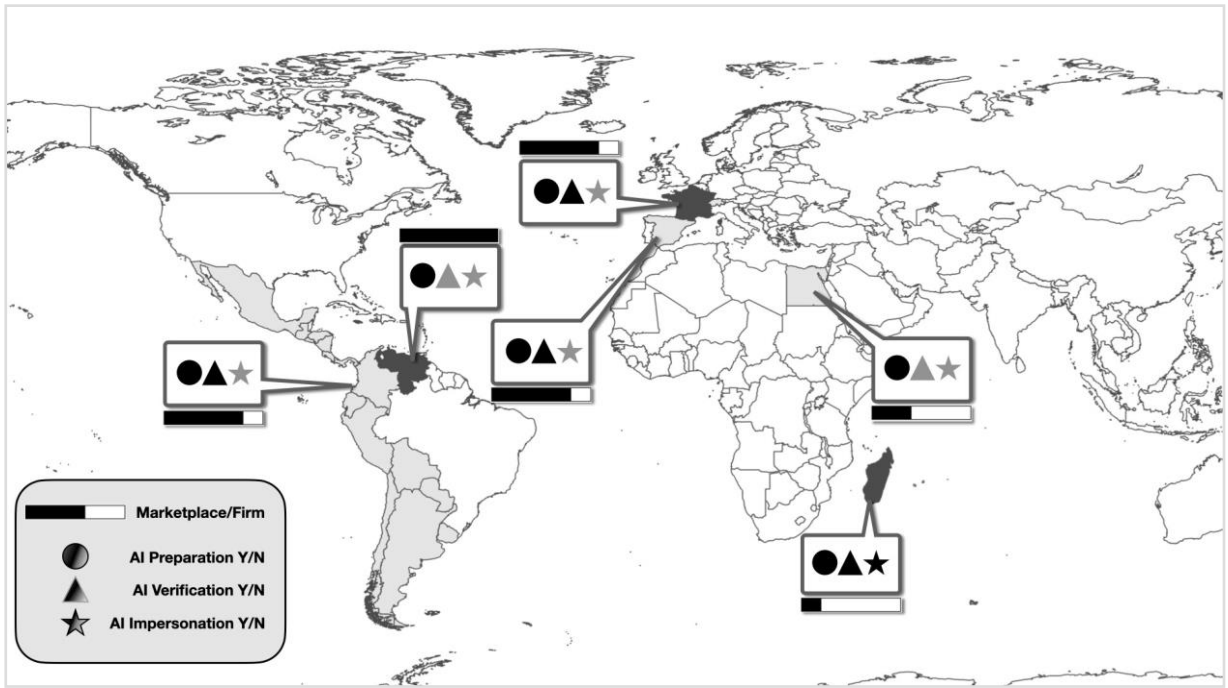